\documentstyle[aps,prl]{revtex}

\author{Ludger Hannibal}
\title{On Hegerfeldt's paradox
}
\date{November 7, 1995
}
\address{
Fachbereich Physik, Carl v. Ossietzky
Universit\"at Oldenburg,
D-26111 Oldenburg, Germany\\e-mail: hannibal@caesar.physik.uni-oldenburg.de}

\begin{document}
\widetext
\maketitle
\begin{abstract}
 The acausal behavior of relativistic states exhibited by Hegerfeldt is shown
not to be present in physical systems
described by first order in time evolution equations.
\end{abstract}



The question of whether relativistic quantum theory is causal or not cannot
be considered settled, as a recent controversy between Hegerfeldt \cite
{Hegerfeldt94} and Buchholz and Yngvason \cite{Buchholz94} shows. In
extension of earlier results \cite{Hegerfeldt74,Hegerfeldt80} Hegerfeldt
\cite{Hegerfeldt94} considered Fermi's \cite{Fermi} two-atom system, where
an atom $A$ is in an excited state, and an atom $B$, which is separated from
$A$ by some distance $R$, is in its ground state and one looks at which time
an excitation of $B$ may occur. Hegerfeldt gave mathematical proof, that
even if the state of $A$ is localized at $t=0$, an excitation of $B$ occurs
with finite probability at any time $t>0$, not only for times $t>R/c$, as one
expects from Einstein causality, i.e. that signal propagation is limited by
the velocity of light $c$. This violation of causality, originally proven
for solutions of the Klein-Gordon equation \cite{Hegerfeldt74}, we call
''Hegerfeldt's paradox''. In their reply Buchholz and Yngvason \cite
{Buchholz94} argued on the general grounds of algebraic quantum field theory
(AQFT) that there no paradox with causality exists, and questioned the
assumptions on localization Hegerfeldt employed. Hegerfeldt takes the
standpoint that his framework goes beyond that of Buchholz and Yngvason, and
that the restrictions inherent to the framework of AQFT render it
inapplicable to the Fermi two atom system \cite{Hegerfeldt95}. Hence the
question raised is whether quantum field theory in its axiomatic \cite%
{Wightman} and algebraic \cite{Haag} foundations, where causality is
well-established, is a sufficient basis for applications in interacting
physical systems. In this short article we want to comment on the restrictions
inherent to both lines of argumentation.

We first consider classical relativistic fields. The initial value problem
for the Klein-Gordon equation
\begin{equation}
\label{1}\left[ \frac{\partial ^2}{\partial t^2}-\sum_{i=1}^3\frac{\partial
^2}{\partial x^{i2}}+m^2\right] \Phi \left( t,{\bf x}\right) =0,
\end{equation}
with Cauchy data given at $t=t_0$,
is solved by (notation: $x=(t,{\bf x}),{\bf %
x}=(x^1,x^2,x^3)$\ , $c=1$)
\begin{equation}
\label{2}\Phi (x^{\prime })=\int \frac{d^4x}{\sqrt{8\pi }}\Delta
(x-x^{\prime };m^2)\left( \delta ^{\prime }(t-t_0)\Phi (t_0,x)+\delta
(t-t_0)\partial _t\Phi (t_0,{\bf x})\right)
\end{equation}
where $\delta$ is the Dirac delta distribution and
\begin{equation}
\label{3a}\Delta (x;m^2)=\Delta _{+}(x;m^2)-\Delta _{+}(-x;m^2)
\end{equation}
with

\begin{equation}
\label{3}\Delta _{+}(x;m^2)=\frac i{2\left( 2\pi \right) ^3}\int e^{-itp^0+i%
{\bf x}\cdot {\bf p}}\frac{d^3p}{p^0},\quad p^0=\sqrt{{\bf p}^2+m^2}
\end{equation}
is the causal propagator for the Klein-Gordon equation. $\Delta$
 is it a tempered
distribution \cite{ReedSimon} (Th. IX.47), with  support
 contained in the forward and backward lightcones \cite{ReedSimon} (Th.
IX.48).

We see from (\ref{2}) that the solutions of the Klein-Gordon equation behave
causally with respect to the joint support of $\Phi (t_0,{\bf x})$ {\it and}
$\partial _t\Phi (t_0,{\bf x})$. This is an induction phenomenon
well-understood in Maxwell theory. In absence of matter, all components of
the electric and magnetic fields satisfy the wave equation (\ref{1}) for $%
m=0 $. The time derivatives of the electromagnetic field components are
interrelated through the Maxwell equations, and (\ref{2}) exhibits the
induction of an electric field by a magnetic field, and vice versa. Similar,
the first order Dirac equation provides a relation between the spinor
components and their time derivatives, while all spinor components satisfy
the Klein-Gordon equation (\ref{1}). In general, if a multicomponent state $%
\Psi $ is governed by a Hamiltonian $H$,%
$$
i\partial _t\Psi =H\Psi
$$
and the Hamiltonian is local in the sense that
$$
supp( H\Psi) \subset supp\; \Psi
$$
then the time evolution is causal with respect to the joint support of all
components of $\Psi $; in this case Hegerfeldt's considerations do not apply.

The basis of Hegerfeldt's arguments is the nonlocal nature of the
Hamiltonian $\sqrt{{\bf p}^2+m^2}$ for scalar fields when spectral
positivity is assumed. Then we cannot have compact supports both of $\Phi
(t_0,{\bf x})$ and$\partial _t\Phi (t_0,{\bf x})$. If $\Phi (t_0,{\bf x})$
has compact support, then it has an analytic three-dimensional Fourier
transform, $\hat \Phi ({\bf p})$, but since $\sqrt{p^2+m^2}\hat \Phi ({\bf p}%
)$, the Fourier of $\partial _t\Phi (t_0,{\bf x})$ with spectral positivity
assumed, is not analytic, $\partial _t\Phi (t_0,{\bf x})$ cannot have
compact support. Looking at the support of $\Phi (t_0,{\bf x})$ alone,
positive frequency solutions appear to behave acausal \cite{Hegerfeldt74}.
One implication is, that positive frequency solutions of first order
relativistic wave equation equations never have compact support \cite
{Thaller}: Hegerfeldt's results cannot be applied to a system of coupled
Fermions and photons as is Fermi's two atom system. Moreover we see a
problem to  interpret the scalar case consistently. If $\Phi ^{*}\partial
_t\Phi -\Phi \partial _t\Phi ^{*}$ is interpreted as charge density, with
the notion that a particle can be located by measurement only where this
quantity is nonzero, we get into conflict with interpreting $\mid \partial
_t\Phi \mid ^2$ as energy density since this latter quantity is not located
where the charge density is; and we cannot speak of energy inducing charge.
In the case of uncharged particles the quantity $\Phi ^{*}\partial _t\Phi
-\Phi \partial _t\Phi ^{*}$ cannot be interpreted as probability density,
since a particle can be located only through its response to external
forces. Since the problems pertain to interacting theories \cite
{Hegerfeldt80} we conclude that a classical theory to be consistent and
causal must not have charged scalar fields. It is interesting to note that
in the framework of general quantum fields theory \cite{Haag} a
corresponding problem
related to causality and the charge structure of a free field theory is well
known. As Haag points out \cite{Haag} (end of Sec. III), the current cannot
be defined as local observable even in free Dirac theory.

In a free quantum field theory the solutions of the classical field
equations form the single-particle Hilbert space. Hence the support
properties are identical and the Hamiltonian of a scalar field is not a
local operator in the sense used above. In the axiomatic framework, the
notion of locality is always related to the support of the test functions,
but this support is not related the support of the Hilbert space states in
configuration representation. Test functions are Fourier transformed and
projected onto mass shells, whereby the compact support in general is lost for
the inverse Fourier-transformed state.

Where Hegerfeldt concludes that the theory has a problem with causality, we
see already from the classical theory that the support properties of Fermion
and photon states will not allow for the existence of space-like separated,
localized states needed for the conclusion.  Buchholz and
Yngvason \cite{Buchholz94}  pointed out that no
 such states with finite energy exist in
algebraic quantum field theory. The theory is fully causal, but may not
contain the desired local observables or localized states, which is
precisely Hegerfeldt's point of criticism \cite{Hegerfeldt95}.

With respect to Fermi's two atom system we conclude, that the assumption of
spectral positivity leads to the result, that in any experimental
preparation the state of an excited atom A will always have a finite overlap
to the state of the non-excited atom B, which allows for causal excitation
without time delay. For experimental purposes, this overlap is relevant only
for distances comparable to the Compton wave lengths of the constituents,
since these are the characteristic lengths of the exponential tails of the
states \cite{Kosinski}.

\end{document}